\documentclass[12pt,preprint]{aastex}
\shorttitle{Age Effects in Ellipticals from Self-Consistent
Simulations} \shortauthors{S\'aiz et al.}

\begin{document}

\title{Ellipticals at $z=0$ from Self-Consistent Hydrodynamical
Simulations: Clues on Age Effects in their Stellar Populations}

\author{ R. Dom\'{\i}nguez-Tenreiro\altaffilmark{1}, A.
S\'aiz$^{1,2}$ and A. Serna$^{3}$}

\altaffiltext{1}{Dept.\ F\'{\i}sica Te\'orica C-XI, Universidad
Aut\'onoma de Madrid, E-28049 Cantoblanco, Madrid, Spain; e-mail:
rosa.dominguez@uam.es. $^2$Current address: Dept. of Physics,
Chulalongkorn University, Bangkok 10330, Thailand; e-mail:
alex@astro.phys.sc.chula.ac.th. $^3$Dept.\ F\'{\i}sica y A.C.,
Universidad Miguel Hern\'andez, E-03206 Elche, Spain; e-mail:
arturo.serna@umh.es}

\begin{abstract}

We present results of a study of the stellar age distributions in
the sample of elliptical-like objects (ELOs) identified at $z=0$
in four hydrodynamical, self-consistent simulations operating in
the context of a concordance cosmological model. The simulations
show that the formation of most stars in each ELO of the sample is
a consequence of violent dynamical events, either fast multiclump
collapse at high $z$s, or mergers at lower $z$s. This second way
can explain the age spread as well as the dynamical peculiarities
observed in some ellipticals, but its relative weight is never
dominant and decreases as the ELO mass at the halo scale, $M_{\rm
vir}$, increases, to such an extent that some recent mergers
contributing an important fraction to the total ELO mass can
possibly contribute only a small fraction of new born stars. More
massive objects have older means and narrower spreads in their
stellar age distributions than less massive ones. The ELO sample
shows also a tight correlation between $M_{\rm vir}$ and the
central stellar l.o.s.\ velocity dispersion, $\sigma_{\rm los,
0}^{\rm star}$. This gives a trend of the means and spreads of ELO
stellar populations with $\sigma_{\rm los, 0}^{\rm star}$ that is
consistent, even quantitatively, with the age effects
observationally detected in the stellar populations of elliptical
galaxies. Therefore, these effects can be explained as the
observational manifestation of the intrinsic correlations found in
the ELO sample between $M_{\rm vir}$ and the properties of the
stellar age distribution, on the one hand, and $M_{\rm vir}$ and
$\sigma_{\rm los, 0}^{\rm star}$, on the other hand. These
correlations hint, for the first time, at a possible way to
reconcile age effects in ellipticals, and, particularly, the
increase of $\alpha $/$<$Fe$>$ ratios with $\sigma_{\rm los,
0}^{\rm star}$, with the hierarchical clustering paradigm. We
briefly discuss the origin of the intrinsic correlations shown by
ELOs in terms of the adhesion model.
\end{abstract}


\section{INTRODUCTION}
\label{intro}

Understanding how galaxies form is one of the most challenging
open problems in cosmology. Elliptical (E) galaxies are the
simplest ones and those that show the most precise regularities in
their measured properties, with their stellar central l.o.s.\
velocity dispersions, $\sigma_{\rm los, 0}^{\rm star}$ (a mass
indicator), strongly correlated with many other of their
properties including luminosities, sizes, colors and index-line
strengths, as recently confirmed by the Sloan Digital Sky Survey
'SDSS' \citep{York00,Bernardi03a,Bernardi03b,Bernardi03c}

The age distribution of stellar populations in E galaxies is one
of the key pieces in the puzzle of the origins and evolution of
galaxies. It was conventionally thought that these populations are
coeval systems, formed as the result of a unique burst of star
formation (SFB) at very high $z$. Recently, spectral indices have
been identified (H$_{\beta}$, H$_{\gamma}$, H$_{\delta}$) that
break the age-metallicity degeneracy, allowing for an improved
stellar age determination in E galaxies through evolutionary
synthesis models \citep[see review in][]{Maraston03}. Even if
still hampered by uncertainties, these models point now to more
massive Es having older mean ages (MAs) and lower rates of recent
star formation and, also, higher suprasolar $\alpha $/$<$Fe$>$
ratios than less massive ones
\citep[e.g.][]{Jorgensen97,Trager00b,Thomas02,Terlevich02,Caldwell03}.
The high $\alpha $/$<$Fe$>$ ratios are actually depressed iron
abundances at higher $\sigma_{\rm los, 0}^{\rm star}$, rather than
$\alpha$ element enhancements \citep{Worthey92,Trager00a}, so that
the $\alpha $/$<$Fe$>$ ratios are a good measure of the timescale
for stellar formation \citep{Weiss95,Thomas99,Pagel01}. The values
of these ratios and their correlation with mass indicators suggest
that an important fraction of the stars in most E galaxies formed
on short timescales, and that this fraction increases with
$\sigma_{\rm los, 0}^{\rm star}$. These trends have been confirmed
by \citet{Bernardi03c} in their analyses of the SDSS sample of E
galaxies, containing to date 9000 galaxies from different
environments. \citet{Jimenez04} reach similar conclusions on the
stellar mass - age correlation through a novel statistical
analysis of $\sim 10^5$ galaxies from SDSS DR1.

Two views have historically existed on how Es formed. The modern
version of the classical {\it monolithic collapse scenario} puts
the stress on E assembly out of gaseous material (that is, with
dissipation), either in the form of a unique cloud or of many
gaseous clumps, but not out of pre-existing stars, with the
stellar populations forming at high $z$ and on short timescales
relative to spirals \citep{Matteucci03}. The competing {\it
hierarchical scenario} \citep[e.g.][]{Toomre77,Kauffmann96}
propounds that galaxies form hierarchically through successive
non-dissipative, random mergers of subunits (the so-called galaxy
merger tree) over a wide redshift range, in such a way that more
massive ones (that is, Es) form more likely at late time. The age
determinations and the interpretation of the $\alpha $/$<$Fe$>$
ratios outlined above, as well as their correlations with mass
indicators, favor the monolithic collapse scenario. In fact, the
hierarchical scenario tends to predict younger MAs and larger
spreads in the stellar age distributions of more massive Es
\citep{Kauffmann98,ThomasK99}, see \citet{Peebles02} and
\citet{Matteucci03} for details and discussions. But the
monolithic collapse scenario does not recover all the currently
available observations on Es either. Such are, for example, the
wide range in ages their stellar populations span in some cases or
their kinematical and dynamical peculiarities
\citep{Trager00a,Menanteau01,deZeeuw02}, indicating that an
important fraction of present-day Es have recently experienced
merger events.

A very convenient approach to reconcile all the current
observational background on Es within a formation scenario, is to
study galaxy assembly from simple physical principles and in
connection with the global cosmological model through
self-consistent gravo-hydrodynamical simulations
\citep{Navarro94,Tissera97,Thacker00}. Individual galaxy-like
objects (GLOs) naturally appear as an output of these simulations,
only star formation processes need further modelling. These
simulations directly provide the phase space structure and the
stellar age distribution of each individual GLO at each $z$, so
that the stellar formation rate history of each GLO can be
determined. This reverses the observational situation where
spectral information is available and stellar age distributions
must be determined through modelling. Also, the structural and
dynamical parameters characterizing each GLO can be estimated. The
first step in the program of studying the origins of E galaxies
through self-consistent simulations, is to make sure that they
produce elliptical-like object (ELO) samples that have
counterparts in the real local Universe as far as structure and
dynamics is concerned \citep[][ hereafter SDS04]{Saiz04}. The
second step is to show that ELO stellar populations have age
distributions with the same trends as those inferred from
observations and to try to understand how these trends arise.
These are the issues addressed in this Letter.

\section{SIMULATED ELLIPTICALS}
\label{simu}

We have run four hydrodynamical simulations (namely, S14, S16, S17
and S26) in the context of a concordance cosmological model, in
which the normalization parameter has been taken slightly high,
$\sigma_8 = 1.18$, as compared with the average fluctuations of
2dFGRS or SDSS galaxies \citep{Lahav02,Tegmark03} to mimic an
active region of the universe \citep{Evrard90}. Galaxy-like
objects of different morphologies form in these simulations. ELOs
have been identified as those objects having a prominent stellar
spheroidal component with hardly disks at all. It turns out that
the 8 more massive objects identified at $z=0$ in S14, S16 and
S17, and the 4 more massive in S26, fulfill this condition. We
report here on age effects in the stellar populations of this ELO
sample. This is the same sample whose structural and kinematical
properties have been analyzed in SDS04, and found to be consistent
with SDSS data \citep{Bernardi03b} for the S16 and S17 subsamples.
We refer the reader to SDS04 for details on the sample general
properties and to \citet{Serna03} for the simulation technique.
Star formation (SF) processes have been included through a simple
parameterization, similar to that first used by \citet{Katz92},
that transforms cold locally-collapsing gas at kpc scales, denser
than a threshold density, $\rho_{\rm thres} $, into stars at a
rate $d\rho_{\rm star}/dt = c_{\ast} \rho_{\rm gas}/ t_g$, where
$t_g$ is a characteristic time-scale chosen to be equal to the
maximum of the local gas-dynamical time, $t_{dyn} = (4\pi
G\rho_{\rm gas})^{- 1/2}$, and the local cooling time; $c_{\ast}$
is the average star formation efficiency at kpc scales, i.e., the
empirical Kennicutt-Schmidt law \citep{Kennicutt98}. To test the
effects of SF parameterization, S14, S16, and S26 share the same
initial conditions and they differ only in the SF parameters
($c_*$ = 0.1, 0.03 and 0.01, and $\rho_{\rm thres}$ = $6 \times
10^{-25}, 1.8 \times 10^{-24},6 \times 10^{-24}$ g cm$^{-3}$ for
S14, S16 and S26, respectively, that is, SF becomes increasingly
more difficult from S14 to S26). To test cosmic variance, S17 is
identical to S16, except that their initial conditions differ.
Supernova feedback effects or energy inputs other than
gravitational have not been {\it explicitly} included in these
simulations. The role of discrete stellar energy sources at the
scales resolved in this work is not yet clear. In fact, in the
context of the new sequential multi-scale SF scenarios
\citep[see][ and references therein]{Elmegreen02,Vazquez03}, some
authors argue that stellar energy releases drive the structure of
the ISM only locally at subkiloparsec scales. Also, recent MHD
simulations of self-regulating SNII heating in the ISM at scales
$<$ 250 pc \citep{Sarson03}, indicate that this process produces a
Kennicutt-Schmidt-like law on average \citep[see
also][]{Elmegreen02}. If this were the case, the Kennicutt-Schmidt
law implemented in our code would already {\it implicitly} account
for the effects stellar processes have on the scales our code
resolves as far as SF rates is concerned, so that our ignorance on
subkiloparsec scale stellar processes relative to SF (not to metal
production or diffusion) rates would be contained in the
particular values $\rho_{\rm thres} $ and $c_{\ast}$ take.

We now briefly comment on the physics of ELO assembly as found in
these simulations \citep[see][]{Sierra03}. The highly non-linear
stages of gravitational instability can be described in terms of
the adhesion model \citep{Shandarin89,Vergassola94}, based on
Burgers' equation \citep[][ 1974]{Burgers48}, whose solutions
exhibit time and space scale invariance in a statistical sense in
terms of the so-called {\it coalescence length}, $L_c(t)$, that
grows as time elapses, defining the average scale of coalescence
in the system. Evolution first leads to the formation of a
cellular structure that is quasi self-similar and not homogeneous,
as $L_c(t)$ depends on the particular region $R$ within the
simulation box considered (and, consequently, hereafter it will be
written as $L_c(t, R)$). At a given scale, overdense regions first
expand slower than average, then they turn around and collapse
through fast global compressions, involving the cellular structure
elements they enclose and in particular nodes connected by
filaments, that experience fast head-on fusions. For the massive
ELOs in the sample this happens between $z \sim 6$ and $z \sim
2.5$. These overdense regions act as {\it flow convergence
regions} (FCRs hereafter), whose baryon content defines the
particles that will end up in a bound configuration forming an ELO
at lower $z$. FCRs contain a hierarchy of attraction basins toward
which a fraction of the matter flows feeding the clumps they host.
Another fraction of the matter keeps diffuse. The transformation
of gas particles into stars at high $z$ mainly occurs through the
multiclump collapse ensuing turn around, that takes the clumps
closer and closer along filaments causing them to merge at very
low relative angular momentum and, consequently, without orbital
delay. This results into strong and very fast SFBs at high $z$
that transform most of the available gas in the clumps into stars,
the exact fraction depending on the values of $L_c(t, R)$ at the
FCR and of the SF parameters. The frequency of head-on mergers
decreases after ELO global collapse, notwithstanding that at low
$z$ mass assembly occurs mainly through merger events. In fact,
many ELOs in our sample have experienced at least one major merger
event at $z < $1, but a strong SFB occurs only if enough gas is
still available after the intense thermo-hydrodynamical activity
epoch at high $z$. So, our simulations confirm the triggering of
SFBs by different dynamical events, either at high $z$ (flow
singularity formation) or lower $z$ \citep[mainly merger or
interaction events][]{Tissera00}. The simulations also show that
diffuse gas is heated and expelled at violent events involving
massive objects, and that a fraction of it is lost to the ELO
potential well. It forms an X-ray emitting corona whose mass
content is similar to the stellar mass content bound to massive
ELOs (that is, a $\sim 50$ \% of the initial gaseous mass).

\begin{figure}
\begin{center}
\includegraphics[width=.8\textwidth]{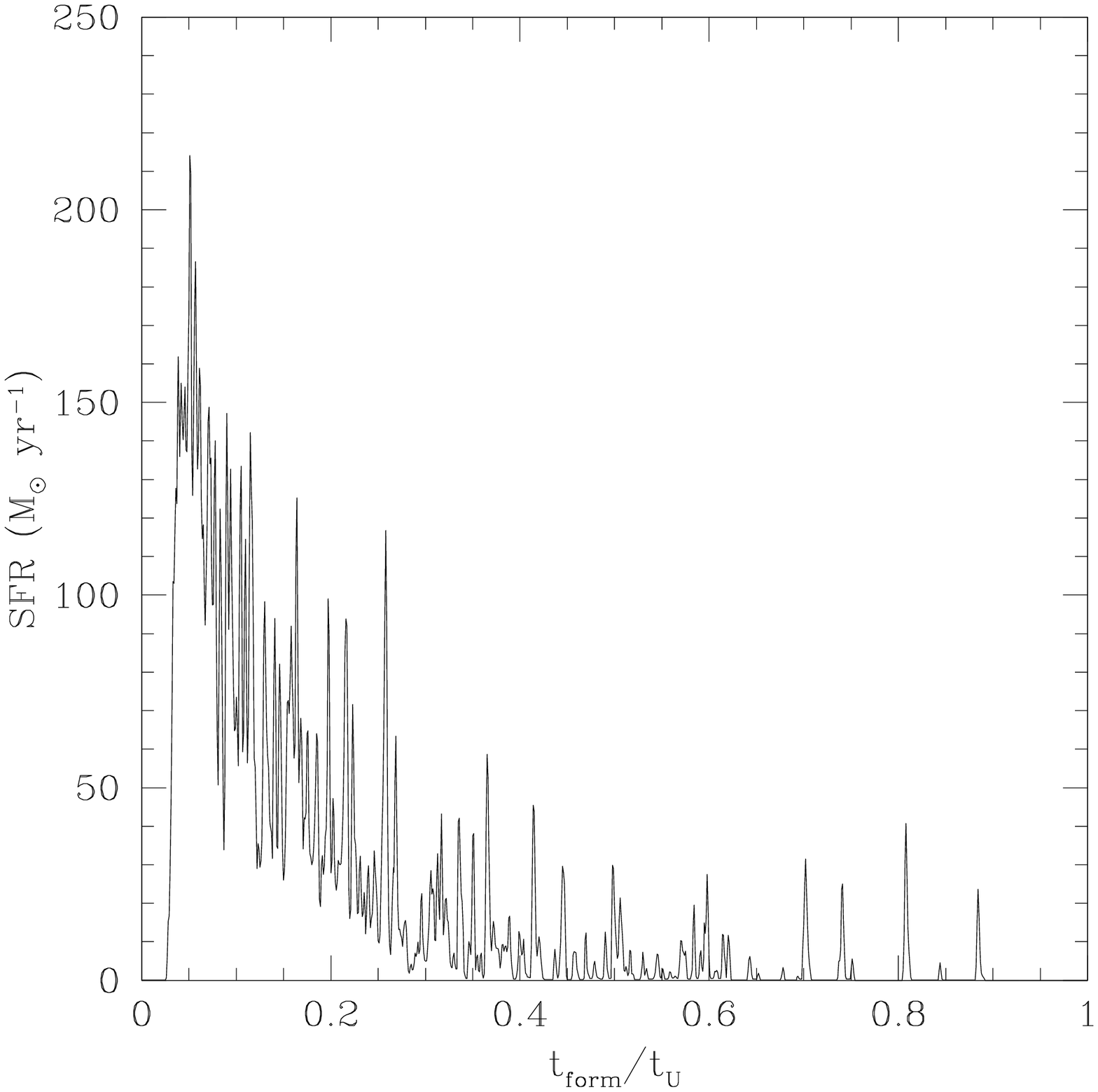}
\caption{Star formation rate history of the more massive ELO
formed in S16 as a function of universe age in units of the
current universe age} \label{sfdif}
\end{center}
\end{figure}

Figure~\ref{sfdif} illustrates the star formation rate history
(SFRH) of a typical massive ELO in the sample (stellar mass
$M_{\rm bo}^{\rm star} = 3.93 \times 10^{11} M_{\odot} $ at
$z=0$). Note that at high $z$ this SFRH is the sum of many SFBs
occurring very close in time. Comparing this SFRH with a detailed
history of the ELO assembly, it has been found out that in fact
the peaks correspond to SFBs in different gaseous clumps toward
which baryons that form the object at $z=0$ flow at early times;
these clumps merge causing or increasing the intensity of their
SFBs, whose width is always narrow. Note that the baryon content
of individual clumps in an ELO FCR is not mixed up before global
ELO collapse. Once collapse is over when the universe age was
$t_{\rm c}$ ($t_{\rm c}/t_{\rm U} \sim 0.3$ in this case), further
SF occurs through random interaction or merger events. An
important parameter is then the mass ratio $M_{\rm bo}^{\rm star,
coll} /M_{\rm bo}^{\rm star}$, giving the fraction of the ELO
total stellar mass at $z = 0$, $M_{\rm bo}^{\rm star}$, formed at
high $z$ in the ELO collapse event. We have found that it
increases with the ELO total mass at the halo scale, $M_{\rm
vir}$, or put in other words, at fixed $\rho_{\rm thres} $ and
$c_{\ast}$, the fraction of gas at the FCR that is transformed
into stars at collapse event increases with $M_{\rm vir}$, leaving
a lower gas fraction available to form stars at lower $z$. At
$z=0$ the cold gas fraction relative to stars inside ELOs is
$\sim$ 2\% to 8\%, depending on their mass. Another important
point illustrated by Figure~\ref{sfdif} is that most SFBs at $t
> t_{\rm c}$ are triggered by minor mergers with small
gas rich clumps, and not necessarily by major merger events, that,
nevertheless are very relevant in ELO mass assembly. In fact, the
ELO in Figure~\ref{sfdif} experienced a major merger (mass ratio
$\sim 0.9$) at $t_{\rm form}/t_{\rm U} \simeq 0.9$ or $z \simeq
0.1$ and, however, no important SFB is apparent in its SFRH,
because in this case negligible amounts of gas bound to the
merging ELOs are available to form stars. We have found that major
mergers of this kind, that is, involving very gas poor objects,
become less frequent as $z$ grows, and, in a given $z$ range, as
$M_{\rm vir}$ decreases. We have also found that more massive ELOs
not only form more stars (an issue already analyzed in SDS04), but
they also form the bulk of their stars earlier on and through
stronger SBs than less massive ones. A consequence is that the age
distribution of stellar populations changes with ELO mass.

\begin{figure}
\begin{center}
\includegraphics[width=.8\textwidth]{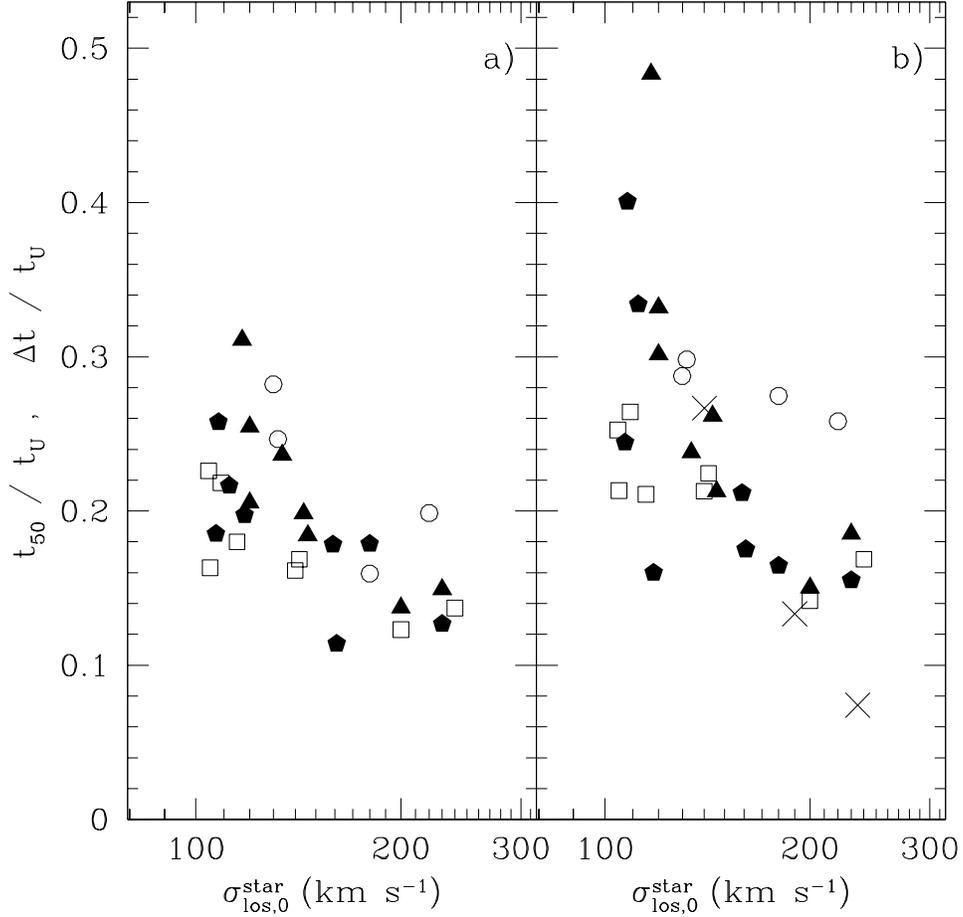}
\caption{(a) Age of the universe in units of the actual universe
age at which the 50 per cent of the total ELO stellar mass at
$z=0$ was already formed, versus their corresponding stellar
central l.o.s. velocity dispersion. Filled triangles and pentagons
stand for S16 and S17 ELOs; open squares and circles for S14 and
S26 ELOs, respectively. (b) Same as (a) for the width of the
stellar population age distribution. Crosses are width estimations
from elliptical data, see text.} \label{Percent}
\end{center}
\end{figure}

To quantify these differences, the percentiles of ELO stellar age
distributions have been calculated. For each ELO in the sample, we
have calculated the redshifts, $z_{f}$, and the universe ages,
$t_{f}$, at which the fraction $f$ of the stellar mass at $z=0$,
$M_{\rm bo}^{\rm star}$, was already formed. The percentiles at
$f=90$ can be taken for a measure of the amount of low $z$ star
formation; the difference $\Delta t = t_{75} - t_{10}$ as an
estimation of the {\it global} width or timescale for ELO star
formation; $t_{50}$ is an estimation of the MA of the population.
For any $f$ a trend exists with $M_{\rm vir}$. The observational
age effects with $\sigma_{\rm los, 0}^{\rm star}$ arise because
$M_{\rm vir}$ and $\sigma_{\rm los, 0}^{\rm star}$ are on their
turn tightly correlated (we have found for the ELO sample
$\sigma_{\rm los, 0}^{\rm star} = (0.057 \pm 0.029) M_{\rm
vir}^{0.28 \pm 0.02}$), so that $\sigma_{\rm los, 0}^{\rm star}$
is an empirical virial mass estimator, see \citet{Saiz03}; S\'aiz
et al.\ (in preparation). Moreover, $\sigma_{\rm los, 0}^{\rm
star}$ and the ELO stellar mass content $M_{\rm bo}^{\rm star}$
are also closely correlated (see details in SDS04), so that a
trend of $t_{f}$ and $z_{f}$ with $M_{\rm bo}^{\rm star}$ also
exists. As an illustration of these trends, in
Figure~\ref{Percent} we plot $t_{50}$ and $\Delta t$ versus
$\sigma_{\rm los, 0}^{\rm star}$ for the ELO sample. We see that
more massive ELOs have older MAs and narrower spreads in the
distributions of their stellar populations. Figure~\ref{Percent}a
compares adequately well with {\it relative} MA determinations
through population synthesis modelling for Es in different
$\sigma_{\rm los, 0}^{\rm star}$ bins. For example, for the
isophotal populations in the E sample from \citet[][ their Tables
1 and 9]{Caldwell03}, the MA averages for Es with $\sigma_{\rm
los, 0}^{\rm star}$ in the ranges (260 - 158) and (158 - 100) km
s$^{-1}$ differ in $\sim 0.1 $t$_{\rm U}$, consistent with that
found for the global populations of the ELO sample
\footnote{\citet{Trager00b} get a difference of $\sim 0.2 - 0.3
$t$_{\rm U}$ for the central stellar populations of the Gonz\'alez
(1993) E sample.}. Figure~\ref{Percent}b shows that width
determinations from $\alpha $/$<$Fe$>$ ratios are consistent with
ELO widths except for the most massive ones, at, say, $\sigma_{\rm
los, 0}^{\rm star} \sim 240 $ km s$^{-1}$ (crosses in
Figure~\ref{Percent}b are from \citet[][ Figure 6]{Thomas02},
maybe because the most massive ELOs have been dynamically shaped
at low $z$ through major merger events involving smaller ELOs of
$\sim$ half their final mass, so that the stellar populations of
these massive ELOs at $z=0$ reflect the properties of the stellar
populations corresponding to the smaller merging ELOs. Note that
the trends in Figures~\ref{Percent} appear in subsamples with very
different SF parameterizations, so that the trends are independent
of the particular details of the SF implementation.

\section{DISCUSSION}
\label{disc}

The simulations we report on in this Letter not only show that ELO
stellar populations have age distributions with the same trends as
those inferred from observations, but they also provide us with
clues on how these trends arise. They indicate that ELOs are
assembled out of the mass elements that at high $z$ are enclosed
by those overdense regions $R$ whose local coalescence length
$L_c(t, R)$ grows much faster than average and whose mass scale
(total mass enclosed by $R$, $M_R$) is of the order of an E galaxy
total (i.e., including its halo) mass. The virial mass of the ELO
at low $z$, $M_{\rm vir}$, is the sum of the masses of the
particles that belong to $R$ and are involved into the ELO merger
tree. Note that this sum is determined by the local realization of
the initial power spectrum of density fluctuations at $R$, that,
on its turn, also determines the total energy of the particles in
the ELO, $E$ (observationally seen as velocity dispersion
$\sigma_{\rm los, 0}^{\rm star}$), and the growth rate of $L_c(t,
R)$. As a consequence, when $L_c(t, R)$ grows faster, $M_{\rm
vir}$ and $E$ are higher (see also S\'aiz et al., in preparation,
and references quoted therein). Now, when $L_c(t, R)$ grows faster
than average at a given region $R$, the local time unit at $R$ is
shorter than at other $R'$s where it grows slower. And so, the
thermo-hydrodynamical activity at high $z$ at $R$ is more intense
than at other $R'$s and, in particular more stars form and gas
becomes exhausted earlier on, making it difficult further star
formation at low $z$ merger events. This is how the intrinsic
correlations age distribution - $M_{\rm vir}$ - $E$ arise in an
ELO sample, whose observational counterpart are likely the age
distribution - $\sigma_{\rm los, 0}^{\rm star}$ correlations shown
by elliptical samples, as discussed in $\S$\ref{intro} and
$\S$\ref{simu}.

The simulations we analyze also suggest that a fraction of the
stars in E galaxies could have been formed at lower redshifts. A
related interesting point is the possibility that mass assembly
and dynamical shaping at low $z$ is not necessarily accompanied by
strong SFB activity. For example, the two more massive ELOs in S14
and S16, or the most massive in S17, have been structurally and
dynamically shaped at a low $z$ ($z \simeq 0.1$ and 0.65,
respectively), however only very modest starbursts resulted from
these dynamically very violent events, because of gas consumption
at high $z$.

Some interpretations of the hierarchical clustering paradigm
consider that the $[\alpha / <$Fe$>]$ - $\sigma_{\rm los, 0}^{\rm
star}$ correlation is not consistent with this paradigm. The
results reported in this Letter indicate that they are indeed,
provided that the bulk of the stellar populations forming ELOs
have been formed in different subunits at high $z$ and merged
together very soon after \citep[see][]{Thomas99,Sierra03}. In this
Letter we show that this fast clumpy collapse follows from simple
physical principles in the context of the current $\Lambda$CDM
scenario.

We are indebted to Dr. M. Moll\'a for useful discussions. This
work was partially supported by the MCyT (Spain) through grants
AYA-0973, AYA-07468-C03-02 and AYA-07468-C03-03 from the PNAyA. We
thank the Centro de Computaci\'on Cient\'{\i}fica (UAM, Spain) for
computing facilities. AS thanks FEDER financial support from UE.

\clearpage

\clearpage


\clearpage

\begin{thebibliography}{}

\bibitem[Bernardi et al.(2003a)]{Bernardi03a} Bernardi, M., et al. 2003a, AJ, 125, 1817
\bibitem[Bernardi et al.(2003b)]{Bernardi03b} Bernardi, M., et al. 2003b, AJ, 125, 1849
\bibitem[Bernardi et al.(2003c)]{Bernardi03c} Bernardi, M., et al. 2003c, AJ, 125, 1882
\bibitem[Burgers(1948)]{Burgers48} Burgers, J. 1948, Adv.\ Appl.\ Mech.\ 1, 171; 1974, "The Nonlinear Diffusion Equation". Reidel Publ., Dordrecht
\bibitem[Caldwell et al.(2003)]{Caldwell03} Caldwell, N., Rose, J.A., \& Concannon, K.D., 2003,
AJ, 125, 2891
\bibitem[de Zeeuw et al.(2002)]{deZeeuw02} de Zeeuw, P.T., et al. 2002, MNRAS, 329, 513
\bibitem[Elmegreen(2002)]{Elmegreen02} Elmegreen, B. 2002, ApJ, 577, 206
\bibitem[Evrard, Silk \& Szalay(1990)]{Evrard90} Evrard, A., Silk, J., \& Szalay, A.S. 1990, ApJ, 365, 13
\bibitem[Jim\'enez et al.(2004)]{Jimenez04} Jim\'enez, R., Panter, B., Heavens, A.F., \& Verde, L. 2004,
astro-ph/0403294 preprint
\bibitem[Jorgensen(1997)]{Jorgensen97} Jorgensen, I. 1997, MNRAS, 288, 161
\bibitem[Katz(1992)]{Katz92} Katz, N. 1992, ApJ, 391, 502
\bibitem[Kauffmann(1996)]{Kauffmann96} Kauffmann, G. 1996, MNRAS, 281, 87
\bibitem[Kauffmann \& Charlot(1998)]{Kauffmann98} Kauffmann, G., \& Charlot, S. 1998, MNRAS, 294, 705
\bibitem[Kennicutt(1998)]{Kennicutt98} Kennicutt, R. 1998, ApJ, 498, 541
\bibitem[Lahav et al.(2002)]{Lahav02} Lahav, O., et al. 2002, MNRAS, 333, 961L
\bibitem[Maraston(2003)]{Maraston03} Maraston, C. 2003, astro-ph/0301419 preprint
\bibitem[Matteucci(2003)]{Matteucci03} Matteucci, F. 2003, Ap\&SS, 284, 539
\bibitem[Menanteau, Abraham \& Ellis(2001)]{Menanteau01} Menanteau, F., Abraham, R. G. {\&} Ellis, R. S 2001, MNRAS, 322, 1
\bibitem[Navarro \& White(1994)]{Navarro94} Navarro, J.F., \& White, S.D.M., 1994, \mnras, 267, 401
\bibitem[Pagel(2001)]{Pagel01} Pagel, B.E.J. 2001, PASP, 113, 137
\bibitem[Peebles(2002)]{Peebles02} Peebles, P.J.E. 2002, in A New Era in Cosmology, ASP Conf., eds.\ N. Metcalf and T. Shanks
\bibitem[S\'aiz, Dom{\'{\i}}nguez-Tenreiro \& Serna(2004)]{Saiz04} S\'aiz, A., Dom{\'{\i}}nguez-Tenreiro, R., \& Serna, A. 2004, ApJ, 601, L131 (SDS04)
\bibitem[S\'aiz(2003)]{Saiz03} S\'aiz, A. 2003, PhD thesis, Universidad Aut\'onoma de Madrid
\bibitem[Sarson et al.(2003)]{Sarson03} Sarson, G.R., Shukurov, A., Nordlund, A., Gudiksen, B., \& Brandenburg, A. 2003, astro-ph/0307013 preprint
\bibitem[Serna, Dom{\'{\i}}nguez-Tenreiro \& S\'aiz(2003)]{Serna03} Serna, A., Dom{\'{\i}}nguez-Tenreiro, R., \& S\'aiz, A. 2003, ApJ, 597, 878
\bibitem[Shandarin \& Zeldovich(1989)]{Shandarin89} Shandarin, S.~F., \& Zeldovich, Y.~B. 1989, Rev.\ Mod.\ Phys.,
61, 185
\bibitem[Sierra-Glez. de Buitrago et al.(2003)]{Sierra03} Sierra-Glez.\ de Buitrago, M.M., Dom{\'{\i}}nguez-Tenreiro, R., \& Serna, A. 2003, in Highlights of Spanish Astronomy III, p.\ 171, eds.\ J. Gallego et al.\ (Kluwer Ac.\ Press)
\bibitem[Tegmark et al.(2003)]{Tegmark03} Tegmark, M., et al. 2003, astro-ph/0310723 preprint
\bibitem[Terlevich \& Forbes(2002)]{Terlevich02} Terlevich, A., \& Forbes, D. 2002, MNRAS, 330, 547
\bibitem[Thacker \& Couchman(2000)]{Thacker00} Thacker, R.J., \& Couchman, H.M.P. 2000, ApJ, 545, 728
\bibitem[Thomas, Greggio, \& Bender(1999)]{Thomas99} Thomas, D., Greggio, L., \& Bender, R. 1999, MNRAS, 302, 537
\bibitem[Thomas \& Kauffmann(1999)]{ThomasK99} Thomas, D., \& Kauffmann, G. 1999, in Spectrophotometric dating of stars and galaxies, ed.\ I. Hubeny et al., Vol.\ 192 (ASP Conf.\ Ser.), p.\ 261
\bibitem[Thomas, Maraston, \& Bender(2002)]{Thomas02} Thomas, D., Maraston, C., \& Bender, R. 2002, R.E. Schielicke (ed.), Reviews in Modern Astronomy, Vol.\ 15, Astronomische Gesellschaft, p.\ 219, astro-ph/0202166 preprint
\bibitem[Tissera(2000)]{Tissera00} Tissera, P.B. 2000, ApJ, 534, 636
\bibitem[Tissera, Lambas, \& Abadi(1997)]{Tissera97} Tissera, P.B., Lambas, D.G., \& Abadi, M.C. 1997, \mnras, 286, 384
\bibitem[Toomre(1977)]{Toomre77} Toomre, A. 1977, in The Evolution of Galaxies and Stellar Populations, eds.\ B. Tinsley \& R. Larson (New Have, CN: Yale Univ.\ Press)
\bibitem[Trager et al.(2000a)]{Trager00a} Trager, S.C., Faber, S.M., Worthey, G., \& Gonz\'alez, J.J. 2000a, AJ, 119, 1645
\bibitem[Trager et al.(2000b)]{Trager00b} Trager, S.C., Faber, S.M., Worthey, G., \& Gonz\'alez, J.J. 2000b, AJ, 120, 165
\bibitem[V\'azquez-Semadeni(2003)]{Vazquez03} V\'azquez-Semadeni, E. 2003, astro-ph/0309717 and
astro-ph/0311064 preprints
\bibitem[Vergassola et al.(1994)]{Vergassola94} Vergassola,~M., Dubrulle,~B., Frisch,~U. \& Noullez,~A. 1994, A\&A, 289, 325
\bibitem[Weiss, Peletier \& Matteucci(1995)]{Weiss95} Weiss, A., Peletier, R.F., \& Matteucci, F. 1995, A\&A, 296, 73
\bibitem[Worthey et al.(1992)]{Worthey92} Worthey, G., Faber, S.M., \& Gonz\'alez, J.J. 1992, ApJ, 398, 69
\bibitem[York et al.(2000)]{York00} York D.G., et al., 2000, AJ, 120, 1579
\end{thebibliography}
\end{document}